\documentclass[%
 reprint, 
superscriptaddress,
 amsmath,amssymb,
 aps,
hyperref
]{revtex4-2}
\usepackage{soul}
\usepackage{physics}
\usepackage{graphicx}% Include figure files
\usepackage{dcolumn}% Align table columns on decimal point
\usepackage{bm}% bold math
\usepackage[colorlinks,citecolor=blue,urlcolor=blue,linkcolor=blue]{hyperref}% add hypertext capabilities

\begin{document}

\title{
Constraints on axion-like ultralight dark matter \\ from observations of the HL Tauri protoplanetary disk}% Force line breaks with \\

\author{Daniil Davydov}
\email[\bf{e-mail:  }]{davydov.dd19@physics.msu.ru} 
\author{Alexander Libanov}%
\email[\bf{e-mail:  }]{libanov.am18@physics.msu.ru}
\affiliation{Institute for Nuclear Research of the Russian Academy of Sciences, 60th October Anniversary Prospect 7a, Moscow 117312, Russia}
\affiliation{%
Faculty of Physics, Lomonosov State University, 1-2 Leninskiye Gory, 119991, Moscow, Russia
}%
\date{\today}
\begin{abstract}
 Dark matter may consist of axion-like particles (ALPs). When polarized electromagnetic radiation passes through the dark-matter media, interaction with background ALPs affects the polarization of photons. The condensate of axionic dark matter experiences periodic oscillations, and the period of the oscillations is of order of years for ultra-light dark matter. This would result in observable periodic changes in the polarization plane, determined by the phases of the ALP field at the Earth and at the source. In this paper, we use recent polarimetric observations of the HL Tauri protoplanetary disk performed in different years to demonstrate the lack of changes of polarization angles, and hence to constrain masses and photon couplings of the hypothetical axion-like ultralight dark matter.
\end{abstract}
\maketitle
\section{Introduction}
\label{intro}
There are many experimental confirmations of the existence of dark matter, but nevertheless, an unambiguous understanding of its nature is still missing \cite{Freese_2017}. One possible explanation is ultra-light dark matter \cite{Marsh:review,Hui:review}, or \cite{PRESKILL1983127, ABBOTT1983133, DINE1983137} that showed that axions could possibly be a dark matter candidate. In the present letter, we assume that ultra-light dark matter (ULDM) consists of axion-like particles (ALPs). This is motivated by the lack of perturbative renormalization of the ALP mass, so that the small mass remains protected from radiative corrections. Dark matter of this type may be detected thanks to the fact that ALPs interact with photons. 

The interaction of axion-like particles with polarized photons emitted by some source affects the polarization phase of these photons. Theoretically, the presence of oscillations of dark-matter axionic condensate is predicted. The period of the oscillations is determined by the ALP mass and, in the case of ultra-light dark matter, is of order of several years. As a result, one could detect a periodic change in the polarization plane of the source \cite{Harari:1992ea,Arvanitaki_2010}. If such oscillations of the polarization plane are not observed, it is possible to constrain the ALP parameters.

Previous works exploited this approach by searching for evidence of oscillations in polarized cosmic microwave background, e.g. \cite{Ferguson_2022,POLARBEAR}, or distant blazars \cite{Ivanov_2019}. It has been pointed out in~\cite{Fujita_2019} that protoplanetary disks represent suitable systems for performing studies of this kind due to their strongly polarized emission, and proximity. They analysed a single observation of the AB~Aur system to establish that the distribution of the positional angle of the polarization plane does not deviate from their astrophysical model predictions. Though they obtained constraints on ULDM ALP parameters,  with only one observation used, the change of the polarization angles with time was not directly constrained, and the conclusions of~\cite{Fujita_2019} depended on the validity of the assumed model.

Here, we take advantage of recently published \cite{stephens2023aligned} ALMA precise polarization map of another protoplanetary disk, HL~Tau, which can be compared to previous observations of the same system with the same instrument, but several years ago \cite{Stephens_2017}. In the rest of the paper, we first remind the reader the theoretical description of the effect of changing the polarization plane of the source during interaction with an oscillating axion condensate of dark matter. Then, we briefly discuss the data we use and the way in which we constrain the change of the polarization angle between two observations. Finally, we obtain our constraints on the ULDM ALP parameters and compare them with others, obtained with similar methods.

\section{ Theoretical description of the expected effect.}
\label{ Theoretical description of the expected effect.}

In this section, we will remind the derivation of the expression for the oscillations of the polarization angle. We follow~\cite{Fujita_2019,Ivanov_2019}. 

Consider the Lagrangian of ALP-photon interaction,
\begin{equation}
    \label{Lagrangian}
    \mathcal{L}=-\frac{1}{4}F_{\mu \nu}^{2} + \frac{1}{2}(\partial_{\mu}a\partial^{\mu}a - m^{2}a^{2}) + \frac{g_{a\gamma}}{4}aF_{\mu\nu}\widetilde{F}^{\mu\nu},
\end{equation}
where $a$ is the ALP field, $m$ is its mass, $F^{\mu\nu}$ is the electromagnetic stress tensor, $\widetilde{ F}_{\mu\nu} = \frac{1}{2}\varepsilon_{\mu\nu\rho\sigma}F^{\rho \sigma}$ is its dual, and $g_{a\gamma}$ is the ALP-photon coupling constant. One can obtain the equation of motion for the electromagnetic field from (\ref{Lagrangian}),
\begin{equation}
    \label{MotionEq}
    \partial_{\mu}F^{\mu\nu}+\frac{1}{2}g_{a\gamma}\varepsilon^{\mu\nu\lambda\rho}\partial_{\mu}(aF_{\lambda\rho}) = 0.
\end{equation}
Using the plane-wave Ansatz for the electromagnetic field,
\[
A_{\nu} = A_{\nu}(k)e^{ikx}+\mbox{h.c.},
\]
we obtain from (\ref{MotionEq}) that the two polarization states propagate with the frequencies
\begin{equation}
    \label{frequency}
    \omega_{\pm} = k \sqrt{1 \pm g_{a\gamma}\frac{\partial_{0}a}{k}} \approx k \pm \frac{1}{2}g_{a\gamma}\partial_{0}a.
\end{equation}
The rotation angle of the polarization plane is therefore given by:
\begin{equation}
    \label{deltaphi}
    \Delta \phi = \frac{1}{2}\int_{t_{1}}^{t_{2}}\!\Delta \omega \, dt = \frac{1}{2}g_{a\gamma}\int_{t_{1}}^{t_{2}} \!\partial_{0}a\,dt,
\end{equation}
where the integration is carried out along the entire path of the electromagnetic wave from the point of emission, $t_{1}$, to the Earth, $t_{2}$.

Assuming $\partial_{0}a\ll\partial_{i}a$, we obtain
\begin{equation}
\label{delta phi main}
\Delta \phi = \frac{1}{2} g_{a\gamma} (a(t_{2}) - a(t_{1})).
\end{equation}

The ALP dark-matter field oscillates as
\begin{equation}
    \label{axionfield}
    a(\Vec{x}, t) = a_{0}(\Vec{x})\cos{(mt+\delta(\Vec{x}))},
\end{equation}
where $a_{0}$ is its amplitude, related to the dark-matter density, and $\delta$ is the phase factor at the location \(\Vec{x}\). This description is valid for times smaller than the coherence time:
\begin{equation}
    \tau_{c} = (m\sigma^{2})^{-1} \approx 2 \times 10^{5} \Big (  \frac{m}{10^{-22}\, eV} \Big)^{-1} \Big (  \frac{\sigma}{10^{-3}} \Big)^{-2} yr,
\end{equation}
where  \(\sigma \sim 10^{-3} \) is the dispersion velocity of dark matter in our Galaxy \cite{Green_2017}.
In our work the relevant time scale is set by the duration of the observations (several years), which is much smaller than the coherence time. This is why the connection between the axionic field amplitude and \(\rho_{DM}\) acquires a stochastic nature. We follow \cite{Castillo_2022}, thus the expression for \(\Delta \phi(t)\) may be rewritten as:
\begin{equation}
    \Delta \phi(t) = \phi_{a} \cos(mt + \varphi_{a}),
\end{equation}
where 
\begin{equation}
    \phi_{a} = 4.48^{\circ} g_{12}  m_{22}^{-1} 
    \Big (  \frac{\rho_{DM}}{1\, GeV cm^{-3}} \Big)^{1/2},
\end{equation}
with $g_{12} = g_{a\gamma}/(10^{-12}\mbox{GeV}^{-1})$, $m_{22} = m/(10^{-22}\mbox{eV})$.
The expression of the phase \(\varphi_{a}\) is not relevant, since it can be considered as a random phase with a flat distribution in \([0,2\pi]\).
 With $\rho_{DM} \approx 0.3\mbox{GeV}/\mbox{cm}^{3}$, the final expression for the change of the polarization angle with time is given by
 \begin{equation}
     \label{delta phi final}
     \Delta \phi \approx 4.3 \times 10^{-2} \cos{\left(mt + \mbox{const}\right)} \, g_{12}m_{22}^{-1}.
 \end{equation}
 
More details can be found e.g.\ in \  \cite{Ivanov_2019,Fujita_2019}; see also \cite{Ivanov1}, or \cite{Castillo_2022}, where the authors developed a new analysis
based on the generalised Lomb-Scargle periodogram to search for periodic signal in the emission of the Crab supernova remnant.

\vskip 1mm

\section{ Data and analysis.}

The protoplanetary disk HL Tau, located at a distance of about 140 pc from the Earth, was observed by ALMA at high resolution. The polarisation maps presented in Fig.~1 of~\cite{Stephens_2017} were obtained with coarse resolution, compared to the most recent work~\cite{stephens2023aligned}. The same plot in \cite{Stephens_2017} reveals that the pattern of polarization angles depends on the wavelength at which the observations were performed. At longer, millimeter, wavelengths it resembles the circular model used in~\cite{Fujita_2019} for AB~Aur, while at the shortest one, 870~$\mu$m, the polarization vectors tend to be parallel to each other across the entire disk. This effect was studied in more detail in the recently published work \cite{stephens2023aligned}. Since the latter presented only 870~$\mu$m results, we restrict ourselves to this wavelength and compare the two observations.

The polarization map presented in~\cite{Stephens_2017} was obtained from observations performed on December 4, 2016. Polarization angles (70 values) were measured on a rectangular grid in the celestial coordinates. The public data accompanying~\cite{stephens2023aligned} is based on a series of observations between June and October, 2021. We split the field of view in rectangular cells, corresponding to the grid used in \cite{Stephens_2017} (but in our grid the center of each cell corresponds to the centre of each vector).
%In each cell, we subtract the polarization angle measured in 2016 from the angles measured in 2021, thus obtaining the distribution of $\Delta\phi$ measured for this cell. 
 Since high resolution corresponds to the length of the red vectors (see~\cite{Stephens_2017} Fig.1, ALMA polarimetric observations at 870µm  )\,, we are interested in the longest vectors. We take the longest 37 vectors, so 37 cells remained (choice of number of vectors is arbitrary, we take about a half of them). For each of these cells, we subtract the polarization angle measured in 2016 from each angle (in the cell) measured in 2021 and add this value to the histogram, thus obtaining the distribution of $\Delta\phi$ measured for this cell. Because each polarisation vector in the 2021 data is not independent from its neighbors, we degrade the grid by taking 1 vector out of 2. Moreover, because the global pattern in the 2021 data is a mix of parallel vectors and of azimuthally oriented vectors, we use weights to take intensity-weighted mean. The normalized errors weights are
 \begin{equation}
\label{restrictions}
  \omega_{i} = \frac{1}{W}\frac{1}{\sigma_{i}^{2}},
\end{equation}
where \(\sigma_{i}  \) are RMS errors, \( W = \sum 1/\sigma_{i}^2\), and with \(\sum \omega_{i} = 1\).

\begin{figure}
    \centering
    \includegraphics[width=1\linewidth]{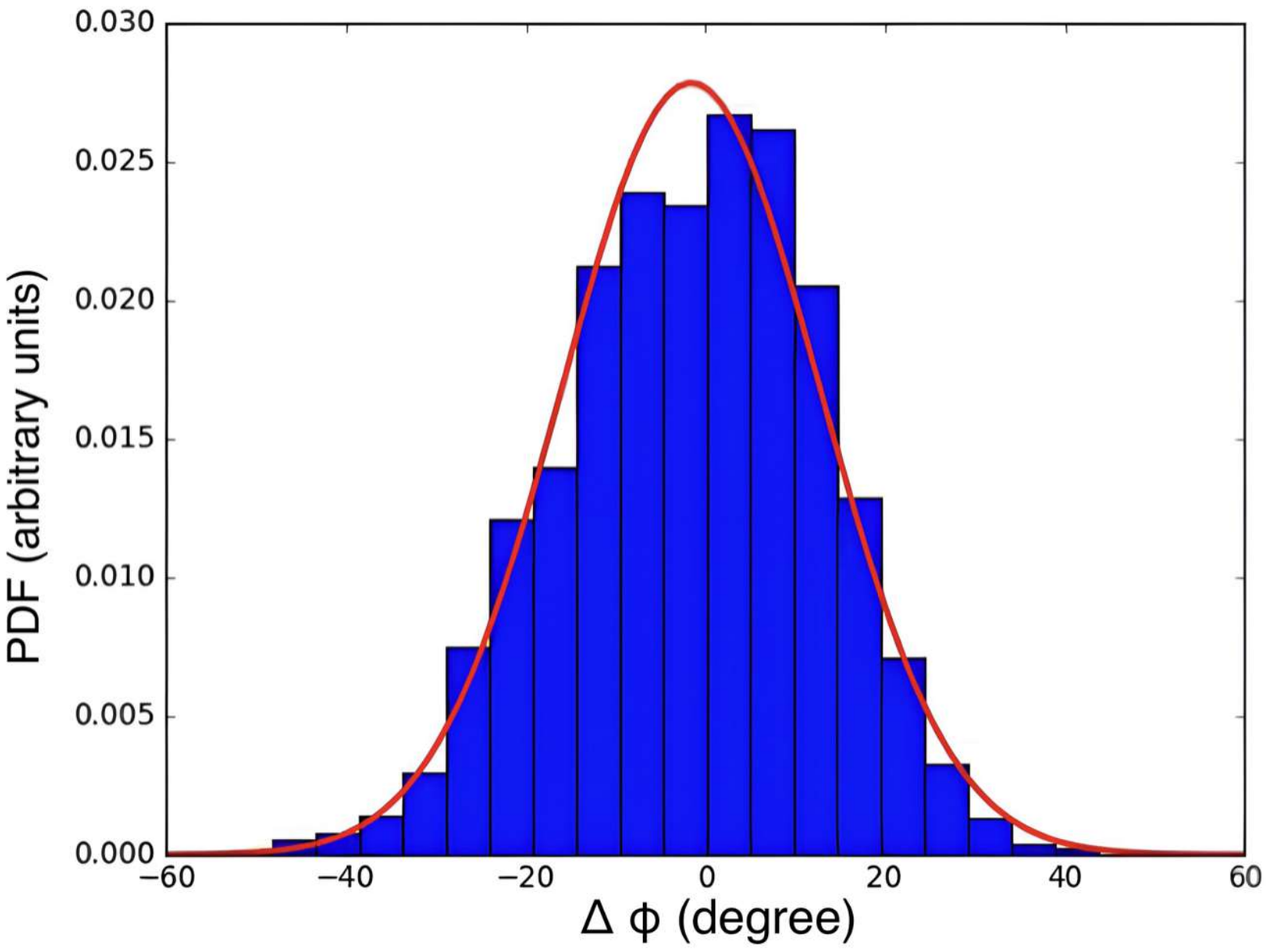}
    \caption{\label{fig:distribution} \sl \textbf{Figure~\ref{fig:distribution}.} Distribution of the polarization angle differences between 2021 and 2016 observations (blue histogram) and the best-fit Gaussian distribution (red line).}
\end{figure}

Finally, we combine all cells to obtain the distribution presented in Fig.~\ref{fig:distribution}. 
We fit this distribution with a Gaussian, keeping both the central value and the width free. The best-fit values of the expectation, $\langle\Delta\phi\rangle  \approx-3.68^\circ$, and the standard deviation, $\sigma_{\Delta\phi} \approx 11.9^\circ$, allow one to constrain 
\begin{equation}
\label{restrictions}
  \Delta \phi < \langle\Delta\phi\rangle \pm 1.96\frac{\sigma_{\Delta\phi}}{\sqrt{N}} \approx 0.0642 \pm 0.0669 ~\mbox{rad}, 95\% \mbox{ C.L.},
\end{equation}

where  $N=37$ is the number of measurements.

\vskip 1mm

\section{ Results.}
\label{ Results.}
By making use of (\ref{delta phi final}) and (\ref{restrictions}), we obtain the constraint 
\begin{equation}
    \label{restrictions 2}
    \cos{\left(mt + \mbox{const}\right)}\,g_{12}m_{22}^{-1} \lesssim 1.49.
\end{equation}

Since we have only two measurements, and one of them is a result of observations during five months, it is hard to use the shape of the cos function here. In addition, the initial phase of the oscillation is unknown. Therefore, following~\cite{Fujita_2019}, we simply replace the cos by $1/\sqrt{2}$ to obtain the final approximate constraint on the ALP-photon coupling constant as a function of the ALP mass, 
\begin{equation}
    \label{restrictions final}
    g_{a\gamma} \lesssim 2.111 \times 10^{-12}~\mbox{GeV}^{-1}\left(\frac{m}{10^{-22}~\mbox{eV}}\right)
\end{equation}

at the 95\% C.L.
\vskip 1mm 

\section{Discussion and conclusions.}
\label{Discussion and conclusions.}
\begin{figure}
    \centering
    \includegraphics[width=1\linewidth]{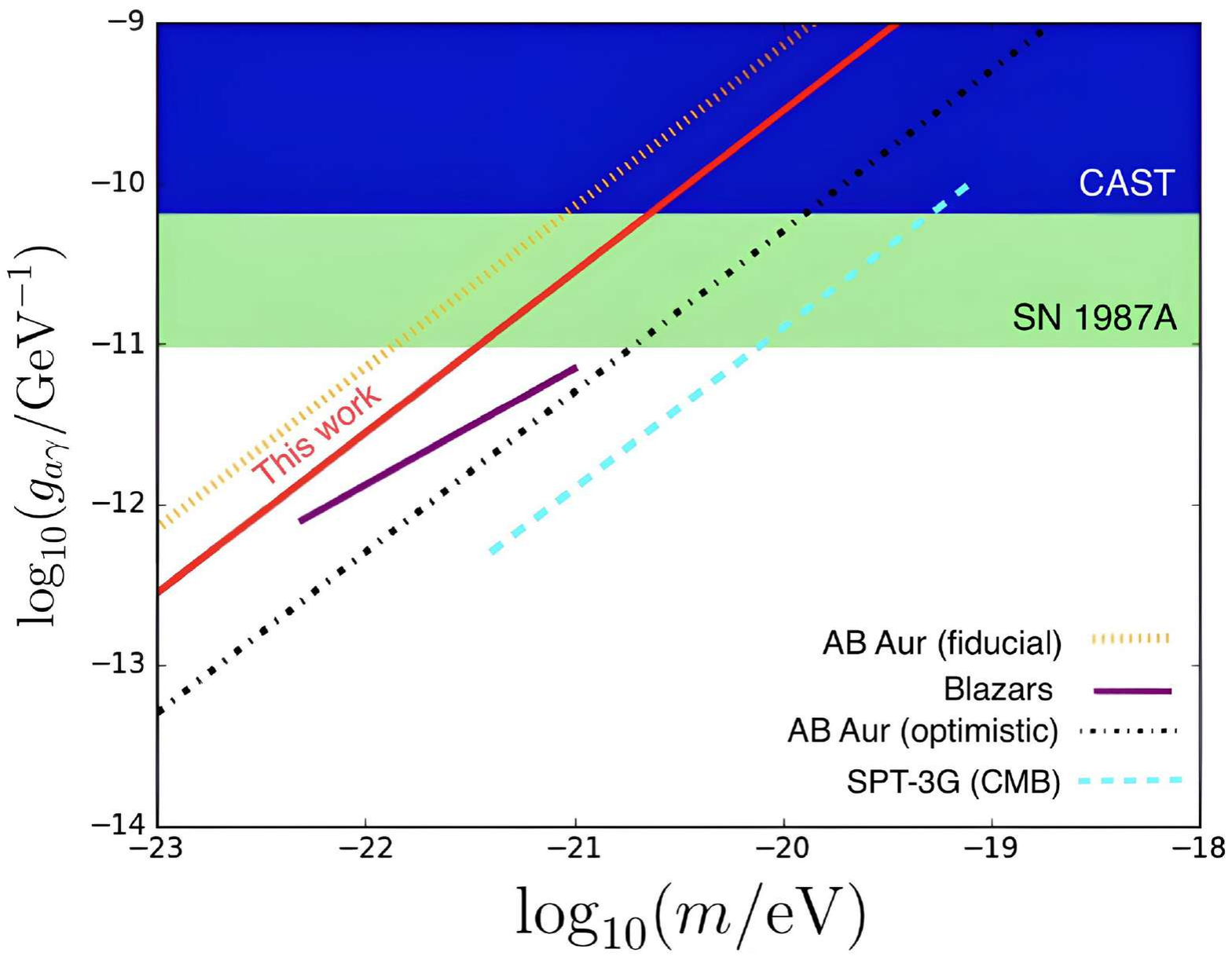}
    \caption{\label{fig:restrictions} \sl \textbf{Figure~\ref{fig:restrictions}.} Comparison of restrictions for constant of interaction between axions and photons $g_{a\gamma}$ and as a function of axion mass. The red line is for restriction from this work. The purple line is for restriction from \cite{Ivanov_2019}. The dashed black and yellow lines is for restriction from \cite{Fujita_2019}. The blue line is for restriction from \cite{Ferguson_2022}.  The green line is for restriction from \cite{Castillo_2022}}
\end{figure}

In Figure \ref{fig:restrictions}, 
we present our constraint compared to others, obtained earlier within similar approaches. We stress that we obtained our limit by direct comparison of two measurements performed with an interval of $\sim 4.5$~years, so that it does not depend on any model assumptions. This is an important step forward from~\cite{Fujita_2019}, though numerically the constraint is of a similar order. The paper~\cite{Fujita_2019} presented two constraints, the ``optimistic'' one, which assumes that their theoretical model describes the physical polarization angle perfectly, and the ``pessimistic'' fiducial one, which attempted to estimate and take into account systematic errors. Our limit is between the two. 

To our best knowledge, only paper~\cite{Ivanov_2019} used long time series to search for, and constrain, the oscillatory pattern in polarization of particular sources. More observations of polarization of nearby and well studied protoplanetary disks, not necessarily as precise as the one presented in~\cite{stephens2023aligned}, but performed at different epochs, would help to improve these constraints.

\begin{acknowledgments}
The authors are grateful to Sergey Troitsky for the original idea and for fruitful discussions on the manuscript. The authors thank Grigory Rubtsov and Maria Kudenko for helpful remarks. This work was supported by the Russian Science Foundation (RSF) grant 22-12-00215.
\end{acknowledgments}
\bibliography{ALP}
\end{document}